# Real-Time Analysis of Nanoscale Dynamics in Membrane Protein Insertion via Single-Molecule Imaging

Chenguang Yang[1,3#], Dongfei Ma[2#], Shuxin Hu[1], Ming Li[1], Ying Lu[1,2,3*].

1 Beijing National Laboratory for Condensed Matter Physics, Institute of Physics, Chinese Academy of Sciences, Beijing 100190, China

2 Songshan Lake Materials Laboratory, Dongguan, Guangdong 523808, China

3 University of Chinese Academy of Sciences, Beijing 100049, China

**Abstract:**

Membrane proteins often need to be inserted into or attached on the cell membrane to perform their functions. Understanding their transmembrane topology and conformational dynamics during insertion is crucial for elucidating their roles. However, it remains challenging to monitor nanoscale changes in insertion depth of individual proteins in membranes. Here, we introduce two single molecule imaging methods, SIFA and LipoFRET, designed for in vitro observation of the nanoscale architecture of membrane proteins within membranes. These methods have demonstrated their efficacy in studying biomolecules interacting with bio-membranes with sub-nanometer precision.

## INTRODUCTION:

Membrane proteins play essential roles in various cellular processes, including ligand-receptor recognition and activation, intercellular communication, and ion transport (Almen *et al.* 2009; Bretscher and Raff 1975; Hegde and Keenan 2022). These proteins often undergo subtle conformational changes to perform their functions. Therefore, it is crucial to dissect the nanoscale architecture and dynamics of membrane proteins when they integrate into the membrane. Over the years, many structural techniques, for instance, X-ray crystallography (Andersson *et al.* 2019; Lieberman *et al.* 2013) and cryo-EM (Cheng 2018; Yao *et al.* 2020), have offered valuable insights into the structures of membrane-interacting macromolecules. However, capturing the real-time conformational changes of membrane proteins continues to be a significant challenge.

Herein, we introduce two single-molecule methods for directly observing the nanoscale dynamics of membrane proteins interacting with bio-membranes, surface-induced fluorescence attenuation (SIFA) (Li *et al.* 2016; Ma *et al.* 2019b) and FRET with quenchers in liposomes (LipoFRET) (Ma *et al.* 2019a). SIFA is based on the fluorescence energy transfer between a fluorophore and monolayer graphene oxide (GO) (Hong *et al.* 2012), so that it is a sensitive point-to-plane distance indicator (Fig. 1A, B). For two single-point coupled dipoles, Like FRET (Förster Resonance Energy Transfer), the emission rate follows a $d^{-6}$ dependence. Replacing one dipole with a line of dipoles results in a $d^{-5}$ scaling upon integration. Similarly, for a two-dimensional array of dipoles on a graphene oxide sheet, a $d^{-4}$ scaling is obtained. (Gaudreau *et al.* 2013) Therefore the distance d between fluorophores and monolayer GO, can be calculated according to the $d^{-4}$ scaling

formula:

$$d = d_0 \left( \frac{F/F_0}{1 - F/F_0} \right)^{\frac{1}{4}}$$

, where F and $F_0$ are the fluorescence of fluorophores in the presence and absence of GO, respectively, and $d_0$ is the characteristic distance at which energy transfer efficiency reaches 0.5. Therefore, SIFA is a powerful tool for measuring the orientation and depth of insertion of membrane proteins in supported lipid bilayers (SLBs) produced by direct vesicle fusion on top of a GO layer modified with PEG (Fig. 1C, D). LipoFRET was developed based on liposomes (Fig. 1E), a bio-mimic system with surface curvature and unrestricted membrane fluidity (Ma et al. 2019a). It is based on the principle of the FRET from one donor to multiple acceptors (quenchers) encapsulated in unilamellar liposomes (Fig. 1F, G). Unlike SIFA, LipoFRET does not have a clear calculation formula for distance versus intensity. This is because the distance between quenchers is comparable to $d_0$ in LipoFRET experiments, the quencher solution cannot be treated as a continuous medium. Instead, Monte Carlo simulations can be utilized to calculate the relative intensity-distance relationships. Fluorophores attached to membrane proteins at different penetration depths (or distance) in the liposome lipid bilayer show different intensities based on their energy transfer efficiency (Fig. 1H). SIFA and LipoFRET can be applied, progressively, to investigate membrane proteins in solid-supported lipid bilayers systems and liposome systems. The techniques utilize the energy transfer between fluorophores and quenchers, and are able to track axial movement of a single fluorophore-labeled protein in membrane.

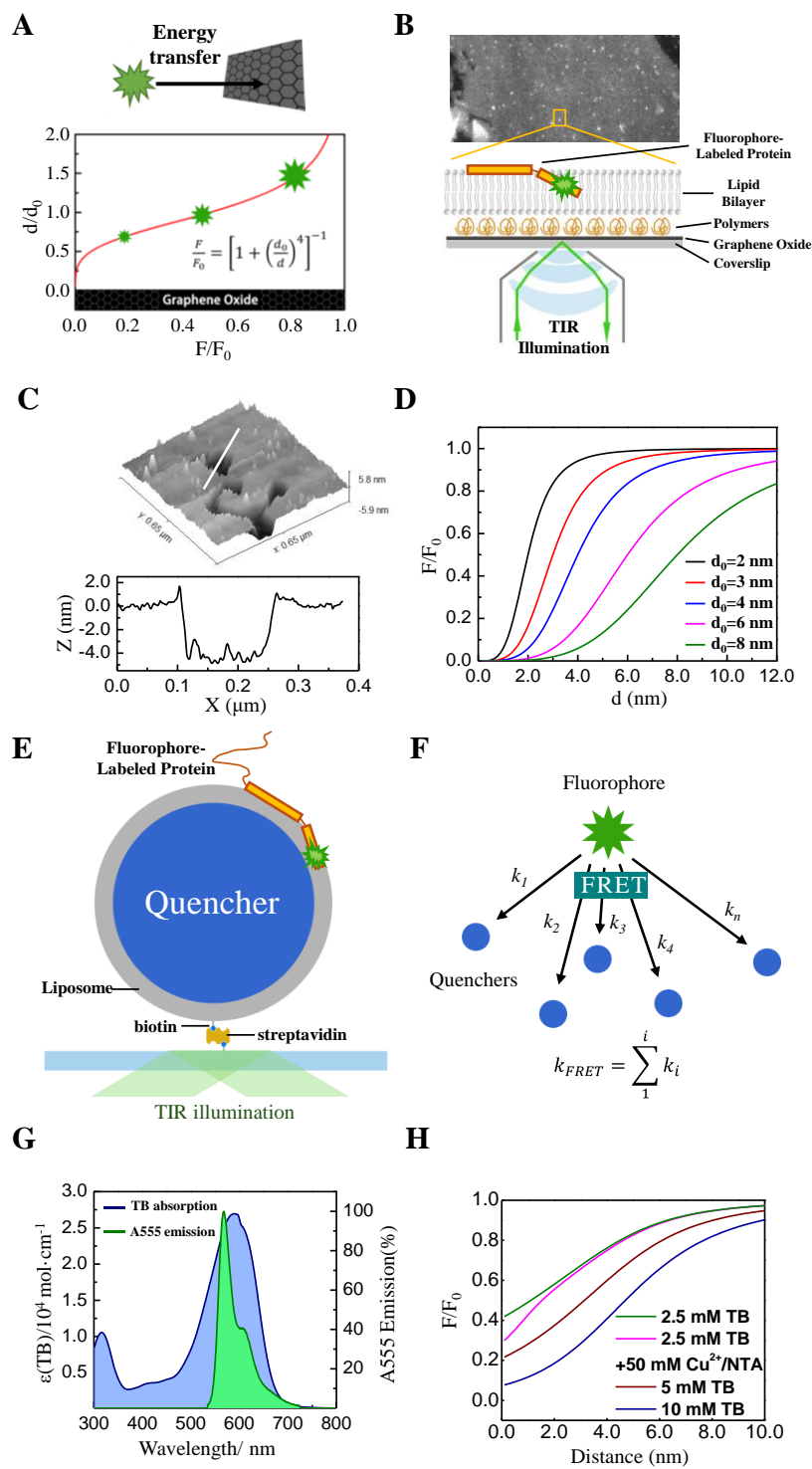

Fig. 1 The principle of SIFA and LipoFRET. (A) The fluorescence of a fluorophore changes rapidly with its distance to a graphene oxide layer. (B) The experimental setup of SIFA on TIRF microscopy. (C) The morphology (upper panel) and the profile of the thickness (lower panel) of a lipid bilayer composed of DOPC/DOPA. (D) The dependence of the relative intensity on distance in SIFA with different critical distances $d_0$. (Ying *et al*. 2016) (E) The scheme of LipoFRET. (F) FRET from one donor to multiple acceptors. (G) The absorption spectra of TB (blue) and the emission spectra of the donor Alexa Fluor 555-MAL (green). (H) The dependence of the relative intensity on the distance to the inner surface of the liposome membrane in LipoFRET. (Ma *et al*. 2021).

**MATERIALS:**

**Biological materials**

- E. coli BL21 strain

**Regents**

- 1-palmitoyl-2-oleoyl-sn-glycero-3-phospho-(1'-rac-glycerol) (POPG; Avanti Polar Lipids)
- 1-palmitoyl-2-oleoyl-glycero-3-phosphocholine (POPC; Avanti Polar Lipids)
- 1-palmitoyl-2-oleoyl-sn-glycero-3-phosphoethanolamine (POPE; Avanti Polar Lipids)
- 1,2-dioleoyl-sn-glycero-3-phospho-(1'-myo-inositol-4',5'-bisphosphate) (PI(4,5)P2; Avanti Polar Lipids)
- 1,2-dioleoyl-sn-glycero-3-phospho-L-serine (DOPS; Avanti Polar Lipids)
- Cardiolipin (CL; Avanti Polar Lipids)
- 1,2-dioleoyl-sn-glycero-3-phosphocholine (DOPC, Avanti Polar Lipids);
- 1,2-dioleoyl-sn-glycero-3-phosphate (sodium salt) (DOPA, Avanti Polar Lipids);
- 1,2-dioleoyl-sn-glycero-3-phosphoethanolamine-N-(cap biotinyl) (Biotinyl Cap PE; Avanti Polar Lipids);

- Trypan Blue (Merck) or Blue dextran, MW 10000 (Merck);
- Chloroform (Fisher Chemical);
- Methanol (Fisher Chemical);
- Acetic acid (Merck);
- 3-(Triethoxysilyl)propylamine (Merck);
- mPEG-SVA, MW 5000 (Laysan Bio);
- Biotin-PEG-SVA, MW 5000 (Laysan Bio);
- Alexa Fluor 555 $C_2$ maleimide (Thermo Fisher)
- PD-10 Desalting Column (Cytiva)
- Superdex™ 75 Increase 10/300 GL (Cytiva)
- Isopropyl β-D-1-thiogalactopyranoside (IPTG; Sigma Aldrich)

**Software and Algorithms**

- MATLAB R2021b (MathWorks)
- Origin 2021 (OriginLab)
- ImageJ 1.52v (National Institutes of Health, USA)

**Equipment**

- Ultrasonic cleaner
- Total internal reflection fluorescent microscope (TIRFM; Nikon Ti2)
- EMCCD IX897 (Andor)
- Oil immersion objective (100x, N.A. 1.49; Nikon)
- QBIS LS 532 nm (CW Solid State Lasers; Coherent)
- Filter setup (49907 ET; Chroma)
- Langmuir-Blodgett (KSV NIMA)
- Extruder kits (Avanti Polar Lipids)

**PROCEDURE**

**1. Procedure of SIFA imaging: Detecting dynamics of membrane protein on the supported lipid bilayers.**

**1.1 Protein expression and purification. [TIMING ~3 days]**

(A) Clone the cDNA that encodes the N-terminal fragments MLKL$_{1–154}$ fused

with a GST tag into the pGEX-4T-2 vector.

**[CAUTION!]** The Alexa Fluor maleimides dyes specifically couple to the thiol groups on Cysteine residue of target proteins. Therefore, the site of target protein for fluorescent labeling must have a single Cysteine residue. To ensure specific labeling, all other Cysteine residues should be mutated. In this manuscript, S55C, S92C and S125C of MLKL$_{1-154}$ is the site that was specifically labelled.

(B) Express the recombinant MLKL in *E.coli* BL21 cells. Grow the cells in Super Broth with 100 µg/mL ampicillin and shake until the OD600 reaches 0.8 at 37 °C.

(C) Protein expression was induced with 0.5 mM IPTG at 18 °C, shaken for a further 16 h.

(D) Harvest cells by centrifugation and lyse by sonication in purification buffer containing 10% glycerol.

**[RECIPE]** Purification buffer: 25mM HEPES, 150 mM NaCl, 0.5 mM TCEP, PH 7.4.

(E) Purify the proteins by using glutathione-sepharose beads and then cleave the GST tag through thrombin.

(F) Purify the proteins using size exclusion chromatography with a Superdex-75 10/300GL size exclusion column.

**1.2 Protein labelling. [TIMING ~15h]**

(A) Mix 10 µM MLKL$_{1-154}$ with 100 µM Alexa Fluor 555-MAL (Alexa Fluor 555 C2 maleimide) dyes at a molar ratio of 1:10 for site-specific fluorescence labeling. The mixture was reacted under 4 °C for 12 hours in the dark.

**[CAUTION!]** Sonicate the dissolved dye in an ultrasonic bath for 15 min to completely dissociate dye oligomers

before adding into target protein, which can avoid cross-linking of dyes. (Zosel *et al.* 2022)

(B) Remove the extra unreacted Alexa Fluor 555-MAL dyes through PD-10 desalting column to and the labelled proteins were flash frozen in liquid nitrogen and stored at – 80 °C until required.

## 1.3 Preparation of LUVs solutions. [TIMING ~10h]

(A) Mix the phospholipids POPC, POPE, DOPS, PI(4,5)P2, CL, and POPG in chloroform at a molar ratio of 35/20/20/10/10/5.

[CAUTIONS!] Chloroform is highly volatile and toxic. Chloroform can dissolve plastics and phospholipids should be mixed in a sealed glass bottle.

[TIP] Note that the composition of the lipids depends on the protein for study.

(B) Dry the chloroform by nitrogen and place the lipids mixture in vacuum pumps over night to remove residual chloroform.

(C) Suspend the dried lipid films in imaging buffer to achieve a final lipid mixture concentration of 1-2 mg/mL. Freeze the liquid in liquid nitrogen, then thaw them in 37℃ water bath. Undergo 10 freezing thawing cycles.

[RECIPE] Imaging buffer: 25mM HEPES, 150 mM NaCl, PH 7.4.

(D) The lipids liquid was forced to go through a polycarbonate filter with 100 nm pore size using a mini-extruder kit for 21 times in order to form large unilamellar vesicles (LUVs) solutions.

## 1.4 Preparation of GO-PEG-supported lipid bilayers. [TIMING ~20h]

(A) Produce ultra-large graphene oxide (GO) flakes by using the modified Hummer's method and collected by centrifugation. (Zhao *et al.* 2010)

(B) Clean the coverslips by

ultrasonicating them in acetone and methanol for 30 minutes each, then treat them with piranha solution at 95°C for 2 hours. After each cleaning step, rinse the coverslips with deionized water and dry them out by nitrogen gas.

[RECIPE] Piranha solution: $H_2SO_4$ (98%) and $H_2O_2$ (30%) at 7:3 volume ratio.

[CAUTIONS!] Piranha solution is highly corrosive, must slowly add $H_2O_2$ to concentrated $H_2SO_4$ while stirring slowly with a glass rod to prevent excessive temperatures and solution splash.

(C) Using the Langmuir-Blodgett to deposit GO monolayer was on surface of the cleaned coverslips and then heat the coverslips under vacuum at 85°C for 2 hours to remove residual solution.

(D) Use double-sided tape to glue the GO-covered coverslip and clean slide into a flow chamber and also glue undeposited GO coverslips and clean slide into chambers for measuring the intrinsic intensity of the fluorophore $F_0$.

(E) Produce a hydrophilic monolayer on the GO-covered coverslips using the cross-linked compound of 1-Aminopyrene and hydroxyl-PEG-NHS ester (AP-PEG). Incubate the coverslips with AP-PEG for 30 minutes, then wash away the unreacted AP-PEG with imaging buffer.

(F) Inject the LUVs solution slowly into the GO-PEG chamber and incubate at 37°C for 12 hours to form GO-PEG-supported lipid bilayers. Wash away extra vesicles with imaging buffer. Repeat these steps for the undeposited GO chamber.

[CAUTIONS!] The chamber should be used within 2 days after the bilayer preparation.

## 1.5 Single molecule SIFA imaging and analysis. [TIMING ~12h]

(A) Mount the GO-PEG chamber onto the stage of the TIRFM.

(B) Seek Large and non-overlapping monolayer GO in the field of view.

[TIP] GO has obvious autofluorescence and the intensity of autofluorescence can be used to determine whether it is a monolayer GO.

(C) Inject ~1 nM labeled MLKL into the imaging GO-PEG chamber and record the fluorescence signal with an EMCCD camera at a frame interval of 30 milliseconds. Output the data as 16-bit TIFF movies for further analysis.

[TIP] Alexa Fluor 555-MAL fluorophores were excited with a 532-nm laser, as were alternative fluorophores like Cy3, ATTO 555, and TAMRA. The filter setup 49907 ET is also suitable for these fluorophores. The frame interval usually varies from 30 ms to 100 ms as required.

(D) Repeat the same imaging steps with undeposited GO chamber for the measurement of the intrinsic intensity of the fluorophore.

(E) Extract the trajectories of fluorophores using a plugin called Particle Tracker in ImageJ and obtain the fluorescence intensity from tiff film based on the trajectory coordinates using a customized MATLAB script. Intrinsic intensity $F_0$ was determined through control measurement and relative intensity ratio $F/F_0$ indicate axial movement of target protein upon lipid bilayers (Figure 2A).

[TIP] To use the Particle Tracker plugin, open your movie and select Particle Tracker from Plugins -> Particle Detector & Tracker. Key parameters include Radius (particle size in pixels, 3 in this

manuscript), Cutoff (score threshold, default 0), Percentile (intensity cutoff for candidate particles, usually 0.2 - 2), Displacement (max pixel movement, default 2) and Link Range (frames for matching, default 10). Adjust and preview as needed.

## 2. Procedure of LipoFRET imaging: Detecting dynamics of membrane protein on the liposome.

### 2.1 Preparation of biotin-carrying liposomes. [TIMING ~12h]

(A) Mix DOPC, DOPA and Biotinyl Cap PE at 7: 3: 0.002 molar ratio and dried into lipid film as protocol described.

(B) Resuspend the lipid film with imaging buffer containing 5 mM quencher (Trypan Blue or Blue dextran). Also resuspend the lipid film with imaging buffer without quencher for the control experiment.

(C) Have lipid mixture solution undergo 10 freezing thawing cycles.

(D) Extrude the lipid mixture as described in protocol. For the liposomes containing quenchers, remove the quencher outside the liposomes with PD-10 desalting columns.

[TIP] In previous studies, liposome size was shown not to affect the intensity-distance curve of LipoFRET. (Ma et al. 2019a) Additionally, the extrusion procedure is capable of producing uniform liposomes, which enable us to adjust the size of the prepared liposomes. This adjustment ensures that curvature of the prepared liposome is suitable for the target membrane proteins to perform their functions.

### 2.2 Coverslip modification. [TIMING ~7.5h]

(A) Clean coverslips as described in protocol till the completion of piranha

solution procedure.

(B) Put the coverslips in the vacuum dryer and let them cool down to room temperature.

(C) Mix the APTES at the volume ratio of 1% in a mixture of methanol and acetic acid (95:5, volume ratio). Incubate coverslips with liquid mixture for 10 min.

(D) Discard the liquid, rinse the coverslips with deionized water and dry them out by nitrogen gas.

(E) Dissolve mPEG-SVA and Biotin-PEG-SVA in the PEG buffer. For 1 mg PEG powder, it would need 10 μL PEG buffer to dissolve it. Mix the mPEG-SVA and Biotin-PEG-SVA at the volume ratio of 99:1.

[RECIPE] PEG buffer: 0.6 M $K_2SO_4$ and 0.1 M $NaHCO_3$.

(F) Add 50-100 μL PEG solution onto one piece of coverslip, then put another piece to stack onto it.

(G) Incubate the coverslips in a humid and dark environment for 2.5 h. Then separate the coverslips, rinse them and dry them out. The coverslips may be stored in vacuum in -20°C.

**2.3 Single-molecule imaging and analysis with LipoFRET. [TIMING ~12h]**

(A) Mount the PEG-coverslip-assembled flow chamber onto the stage of TIRFM.

(B) Add streptavidin at 0.01 mg/mL into the chamber and incubate for 10 min. Wash the chamber with the buffer to remove excess streptavidin.

(C) Dilute the quencher-encapsuled liposomes to ~0.01 mg/mL, inject them into the chamber. Incubate for 5 min. Wash the chamber again with the buffer to remove unbound liposomes.

[TIP] Because the wide adsorption spectra and autofluorescence of the quencher, one may observe the liposomes

in an emission channel that fits Cy5 dye to determine if the sufficient density of the liposomes is reached.

(D) Add the target protein to the channel to interact with the liposomes, and record images or films. In this example, α-synuclein labeled by Alexa Fluor 555-MAL was also obtained by a similar purification and labelling procedure as described in protocol. The labeled α-synuclein (α-synuclein K10C-Alexa 555, α-synuclein T72C-Alexa 555, or α-synuclein S129C-Alexa 555) was added at ~1 nM.

(E) Repeat the above imaging steps with the liposomes that do not contain quencher for the measurement of the intrinsic intensity of the fluorophore.

(F) Extract the intensity of the fluorophores with software ImageJ and MATLAB. A gaussian fitting can be applied to derive the intrinsic intensity of the fluorophore-labeled proteins ($F_0$) without quencher. Then the fluorescence intensity (F) of the proteins on quencher-containing liposomes is also analyzed. The normalized intensity $F/F_0$ and its changes are compared to the intensity-distance curves to derive the position changes of the labeled site during the conformational motion of the protein.

**ANTICIPATED RESULTS**

In this protocol, we describe the procedure of SIFA and LipoFRET in detail. As shown in Figure 2, SIFA enables us to track the three-dimensional movement of MLKL on supported lipid bilayers with an experimental accuracy of ~0.6 nm for axial movement and ~25 nm for lateral movement (Figure 2A). Appling SIFA to different interest site of MLKL (Figure 2B and 2C), two states 'Anchored' and 'Embedded' were captured, even though only nanoscale change in architecture between two states, and nanoscale dynamics of MLKL undergo conformational change were detected. Whereas LipoFRET distinguishes the penetration depths of the proteins on the membrane of liposomes. Application of LipoFRET on α-synuclein could detect the spontaneous intensity alteration of α-synuclein K10C-Alexa 555, which represents the shift among three penetration depths (Figure 2D). The intensity of α-synuclein S129C-Alexa 555 can also be seen higher than that of α-synuclein T72C-Alexa 555. In addition, the C-terminal of the protein and the membrane surface could be observed to move to a closer position relative to the membrane surface upon the $Ca^{2+}$ addition (at 0.1 to 1 μM concentration) when applying the method to α-synuclein S129C-Alexa 555 (Figure 2E).

## PERSPECTIVES

Single-molecule fluorescence imaging has long been applied in membrane protein studies, including techniques like FRET and FIONA (Fluorescence Imaging with One Nanometer Accuracy). FRET has been widely used to probe the folding of membrane proteins, intra-molecular movements of domains, and inter-molecular movements of interacting proteins. (Mercier *et al.* 2020; Shi *et al.* 2024) FRET is suitable for a detection range of approximately 3 to 8 nm. (Ha 2001) FIONA excels in single-molecule localization, providing axial accuracy of 0.5 nm and lateral accuracy of 1-2 nm, offering unparalleled precision in localizing individual molecules and tracking their movements over time. (Park *et al.* 2007; Selvin *et al.* 2007; Wang *et al.* 2014)

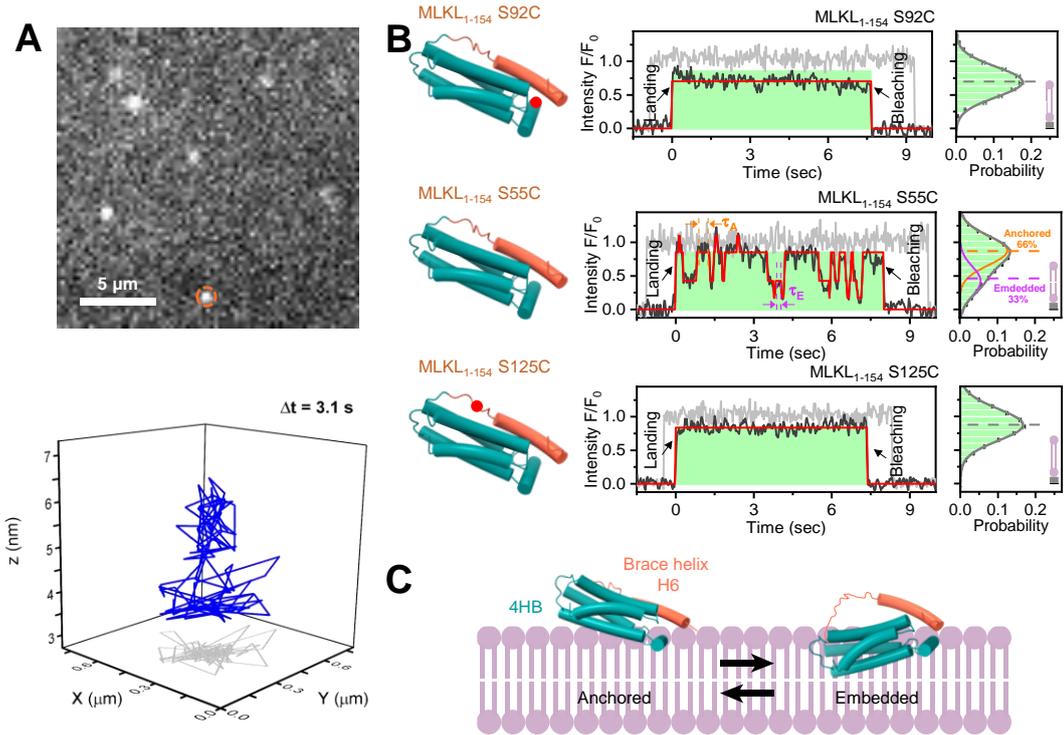
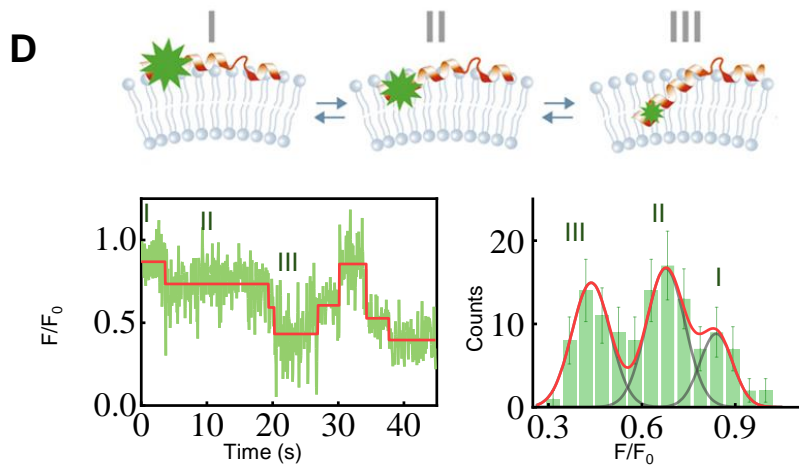
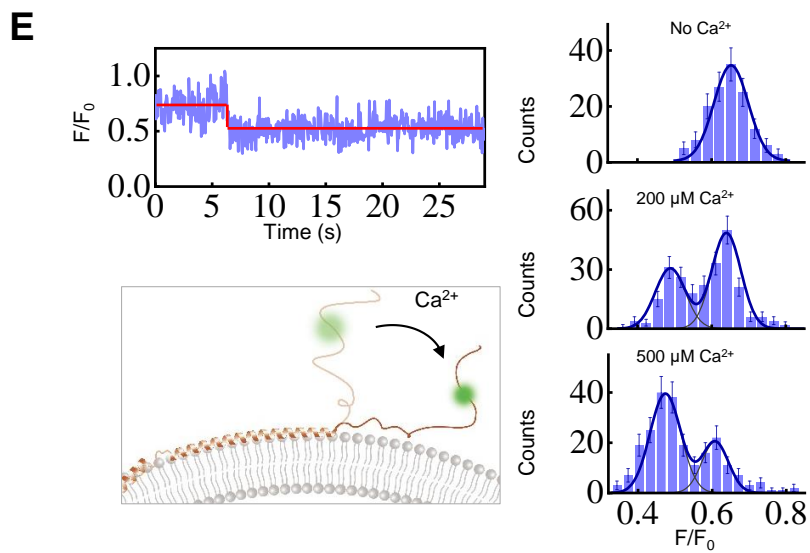

**Fig. 2** Observing the interaction of membrane protein with membranes with SIFA and LipoFRET. **(A)** A 3D trace of a singly labelled MLKL protein. The in-plane positions (the xy coordinates) were read from the image, and the z coordinates were derived from the intensity F(x,y) of the fluorophore. **(B)** Typical fluorescence trajectory (middle) and statistics of intensity (right) of S92C, S55C and S125C labelled MLKL$_{1-154}$ (left), respectively. **(C)** Cartoon representation of the two states. **(D)** The spontaneous membrane insertion dynamics of the N-terminal of α-synuclein (upper panel) reflected by intensity trace (lower left) and the histogram (lower right). **(E)** The position changes of the C-terminal of α-synuclein upon calcium adding (left) and the position distribution characterized by the intensity distribution(right). (Ma *et al.* 2019a; Yang *et al.* 2023)

SIFA and LipoFRET are capable of detecting the insertion of membrane proteins into biological membranes with sub-nanometer experimental accuracy. SIFA is suitable for tracking the three-dimensional movement of target proteins (Jiang *et al.* 2022; Ma *et al.* 2019b; Ma *et al.* 2018; Yang *et al.* 2023), while LipoFRET is ideal for detecting the location of the curvature-sensitive membrane proteins inserted in bio-membranes. (Ma *et al.* 2020; Ma *et al.* 2019a) Each technique has unique strengths, making them suitable for different experimental needs.

**Acknowledgements** This work was supported by GuangDong Basic and Applied Basic Research Foundation (2020A1515110102), the National Natural Science Foundation of China (12022409 and T2221001) and the CAS Key Research Program of Frontier Sciences (ZDBS-LY-SLH015).

**Compliance with Ethical Standards**

**Conflict of interest** Chenguang Yang, Dongfei Ma and Ying Lu declare that they have no conflict of interest.

**Human and animal rights and informed consent** This article does not contain any studies with human or animal subjects performed by any of the authors.